# A Novel Method of Discrete-Time Amplification using NEMS Devices

Sivaneswaran Sankar, Maryam Shojaei Baghini, Valipe Ramgopal Rao
Dept. of Electrical Engineering
Indian Institute of Technology Bombay
Mumbai, India
{sivaneswaran, mshojaei, rrao}@ee.iitb.ac.in

*Abstract*— In this paper we propose a novel method of realizing discrete-time (D-T) signal amplification using Nano-Electro-Mechanical (NEMS) devices. The amplifier uses mechanical devices instead of traditional solid-state circuits. The proposed NEMS-based D-T amplifier provides high gain and operate on a wide dynamic range of signals, consuming only a few micro watts of power. The proposed concept is subsequently verified using Verilog-A model of the NEMS device. Modifications in the proposed amplifier to suit different specifications are also presented.

## I. INTRODUCTION

The development of integrated circuits using mechanical switches has attracted significant interest in the recent years due to its potential in reducing the overall energy consumption of the system [1]-[8]. Many have utilized NEMS switches for digital logic application, owing to its high ratio of resistances in the on-state and off-state. In [9], continuous-time amplification of the signal using variable MEMS capacitor is achieved. NEMS switch as a transconductance element is proposed in [10] for analog applications.

Switched capacitor networks are inherently present in many ADCs, DACs, DC-DC converters and so on. The most common signal processing function needed is the amplification. In D-T systems, amplification of the input signal is traditionally implemented using a capacitive feedback around an operational amplifier (op-amp) [11], as shown in Fig.1. In the sampling phase ($\phi_1$), the input "$V_{in}$" is sampled across capacitor $C_1$. In the hold phase ($\phi_2$), the charge "$C_1V_{in}$" which was present initially in $C_1$ is transferred to the capacitor $C_2$ by the op-amp. Hence, if $C_2<C_1$, one can obtain a voltage gain equal to $C_1/C_2$.

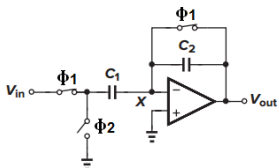

Fig.1. Traditional swiched capacitor amplifier (from [11])

Specifications of the op-amp greatly influence the performance of the D-T amplifier and it is the most power hungry block. Some of the reported works in which D-T amplification was performed without using op-amp are reported in [12]-[15]. In [12] comparator is used while [13] uses dynamic source follower, [14] employs charge pump technique and [15] utilizes MOS varactor. In this work we propose to utilize NEMS devices for performing D-T signal amplification. The proposed amplifier can provide high gain for wide range of signals and can be modified for different gain or voltage specifications.

This paper is organized as follows. In Section II, the terminal characteristics of the NEMS devices are presented. Section III describes the concept of D-T amplification using NEMS. In Section IV, the simulation results are discussed.

## II. PRINCIPLE OF OPERATION OF NEMS SWITCH

In this work, two kinds of NEMS devices are utilized: (1) NEMS capacitive switch (for signal amplification), (2) NEMS ohmic switch (for sampling the signal).

### A. NEMS Capacitive Switch

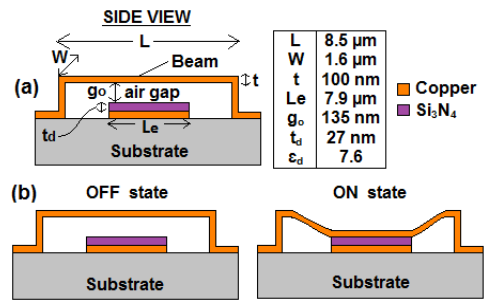

Fig. 2. (a) Schematic of NEMS capacitive switch, (b) Switch in off and on-state

The schematic of the NEMS capacitive switch is shown in Fig. 2(a). The beam clamped at two ends, forms the top plate of the capacitor. Between the top plate and the bottom plate there is a bi-layer consisting of air and a dielectric material ($\varepsilon_r$). If the electrostatic force (due to applied voltage) between the two plates is higher than the spring restoring force of the beam, then the beam gets pulled-in and the switch is set in on-state, else it is off. For the on-condition, $|V|>V_{PI}$ (pull-in voltage) and for the off- condition, $|V|<V_{PO}$ (pull-out voltage). Fig. 2(b) depicts the state of the switch in off and on-state. The C-V curve and transient characteristics of the NEMS capacitive switch modeled using Verilog-A [16] is plotted in Fig. 3.

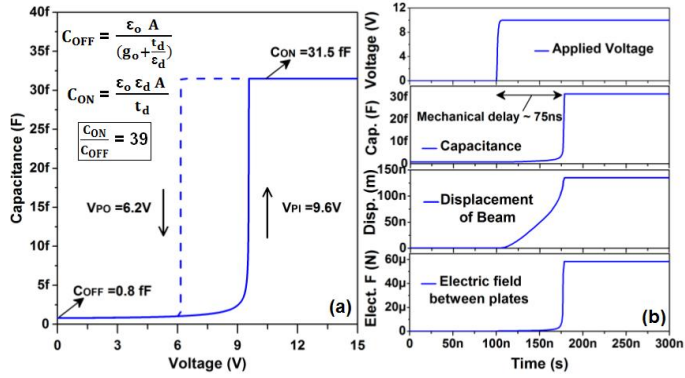

Fig.3. (a) C-V curve, (b) Transient characteristics of NEMS Capacitive switch

### B. NEMS Ohmic Switch

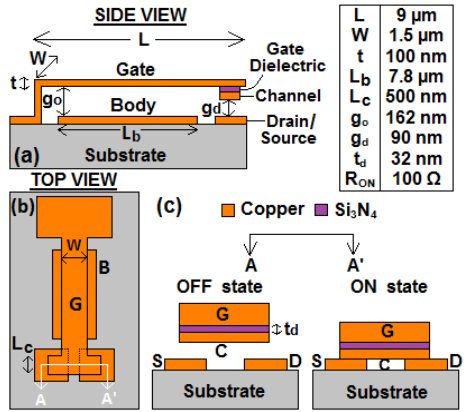

Fig.4. (a) Side view of NEMS Ohmic switch, (b) Top view of the switch, (c) cross section of the switch along A-A' during off and on-state

The side view and top view of the NEMS ohmic switch is shown in Fig.4. (a) and (b) respectively. The cantilever beam fixed at one end forms the gate terminal, to which the floating channel is attached through the dielectric material. The electrostatic attractive force between the gate and body sets the state of the switch. For the on-condition, $|V_{GB}|>V_{PI}$ (pull-in voltage) and for the off- condition, $|V_{GB}|<V_{PO}$ (pull-out voltage). Fig.5. (c) depicts the state of the switch in off and on-condition. The I-V characteristics and the transient characteristics of the NEMS ohmic switch modeled using Verilog-A [16] is plotted in Fig. 5.

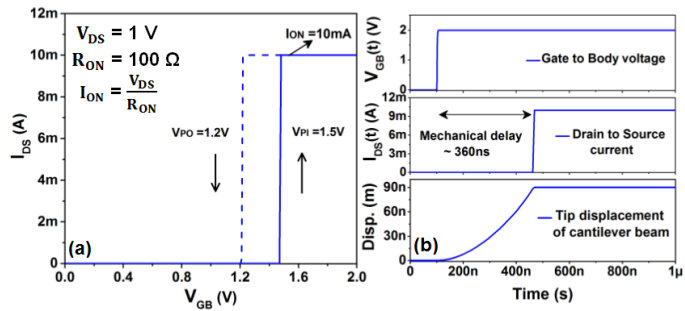

Fig.5. (a) I-V curve, (b) Transient characteristics of NEMS Ohmic switch

## III. CONCEPT OF D-T AMPLIFICATION USING NEMS

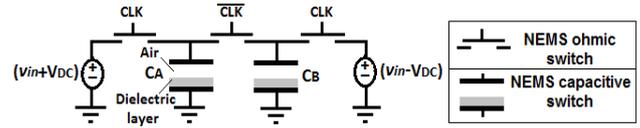

Fig.6. Schematic of the NEMS D-T amplifier. *vin* is the signal that is being amplified (*vin* < $V_{DC}$) and $V_{DC}$ > $V_{PI}$ of NEMS Capacitive switch.

The schematic of the D-T amplifier using NEMS is given in Fig.6, where $C_A$ and $C_B$ represent the NEMS capacitive switch and the three NEMS ohmic switches are controlled by clock signals. "$V_{DC}$" is assumed to be greater than the pull-in voltage of the NEMS capacitive switch and "*vin*" is the signal that is being amplified. The mechanism of amplification is depicted step wise in Fig.7.

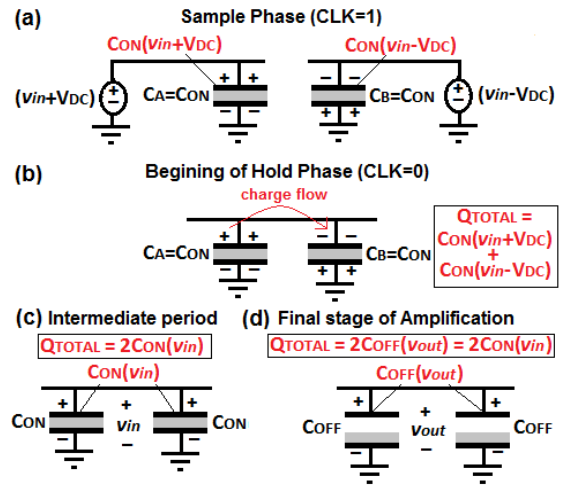

Fig.7. Stages of amplification: (a)Sample Phase $0<t\leq T/2$, (b)Begining of hold Phase $t=(T/2)^+$, (c)Intermediate period $t=(T/2)^{++}$, (d) Final step $T/2 \ll t \leq T$. Charges are denoted in *red color*. T = Time period of Clock.

**Step (a):** During the sample phase (CLK=1) the voltages "*vin*+$V_{DC}$" and "*vin*-$V_{DC}$" are sampled across the two NEMS capacitors as shown in Fig.7. (a). If $|vin \pm V_{DC}| > V_{PI}$, the beams get pulled-in due to the always attractive nature of the electrostatic forces. Hence the voltages are sampled across a larger capacitance "$C_{ON}$".

**Step (b):** Now in the hold phase (CLK=0), moment the two capacitors are shorted as shown in Fig.7. (b), charges flows from $C_A$ (left one) to $C_B$ (right one) due to the difference in potentials across them. The total charge in the top plates of the capacitors are given by, $Q_{TOTAL} = C_{ON}(vin+V_{DC})+C_{ON}(vin-V_{DC})$.

**Step (c):** As evident, the charge due to $V_{DC}$ gets cancelled and $Q_{TOTAL} = 2 \times C_{ON}(vin)$. The voltage across the capacitors is now "*vin*", which is shown in Fig.7. (c) and is assumed to be lesser than the pull-out voltage of the NEMS capacitive switch.

**Step (d):** Since "*vin*"<$V_{PO}$, the beams get released due to the fact that electrostatic force is not strong enough to overcome the spring force of the beam. As a result, the capacitances $C_A$ and $C_B$ reduce to $C_{OFF}$. For the charge conservation to hold, $Q_{TOTAL} = 2 \times C_{OFF}(vout) = 2 \times C_{ON}(vin)$. Hence *vout* gets amplified to $(C_{ON}/C_{OFF}) \times vin$.

## IV. SIMULATION RESULTS OF THE NEMS D-T AMPLIFIER

### A. Waveforms

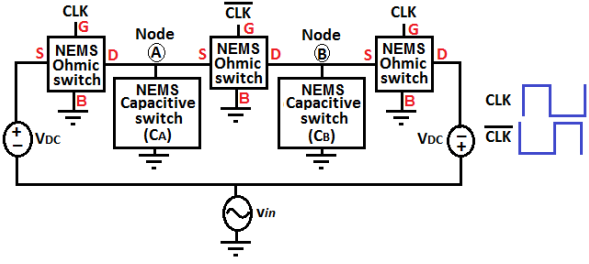

Fig.8. Circuit implementation of the D-T amplifier using Verilog-A model.

The implementation of the D-T amplifier in Cadence using Verilog-A model is shown in Fig.8. Based on the parameters of the NEMS capacitice switch described in Section II, the maximum voltage gain achievable is 39. The clock signals are provided through the non-overlapping clock generator circuit (not shown here). The clock frequency ($f_{CLK}$) and $V_{DC}$ is chosen to be 100 KHz and 10 V (throughout the work). Lets consider "$vin$" to be a DC signal equal to 10mV. The voltages at node A and B are plotted in Fig. 9.

In the sample phase, "10mV+10V" is sampled across $C_A$ and "10mV-10V" is sampled across $C_B$. In the hold phase, the voltage across the node A and B are equal to 389.9 mV, thus providing a gain equal to 38.99. Important internal variables of the NEMS capacitor $C_A$ that is connected to node A is shown in Fig. 10. The circuit in Fig. 8 effectively performs the function of the circuit shown in Fig. 1. For a sinusoidal input signal, the sampled and amplified differential output waveform is plotted in Fig. 11.

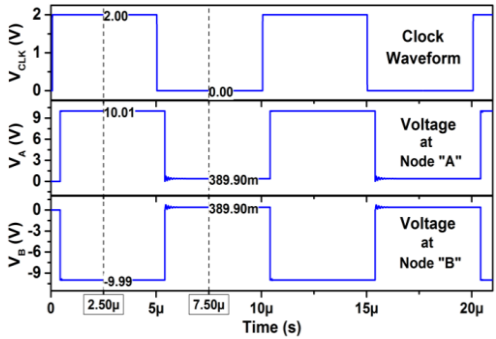

Fig.9. Voltage waveforms at Node "A" and "B". $vin$ =10mV.

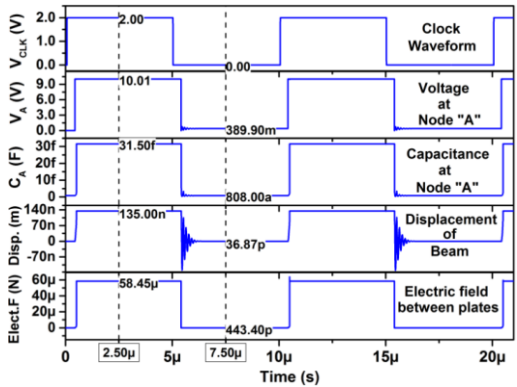

Fig.10. Detailed transient charactistics of the NEMS Capcitive switch "$C_A$".

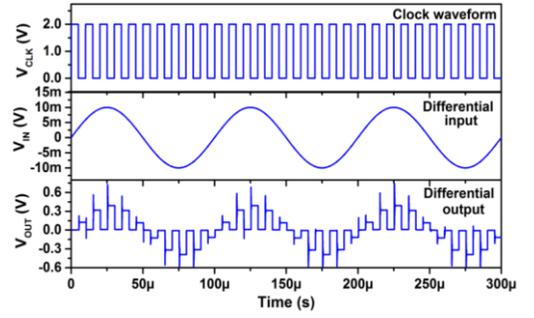

Fig.11. Sample and hold amplification of sinusoidal input using differential NEMS D-T amplifier. Peak differential $vin$ is 10mV at a frequency of 10 KHz.

### B. Non-idealities

One important point to consider is that, the amplified signal "$vout$" will cause the beam to displace and change the value of capacitance in the off-state. From Fig. 3, we can observe that the capacitance in the "off-state" of the NEMS capacitive switch changes with its terminal voltage in a negligible fashion (i.e slope~0) until the pull-out point. Hence it gives rise to weak non-linearity as evident from Fig.12, where the gain (*denoted in blue color*) drops only by 8% for an increase in the input amplitude by 175×.

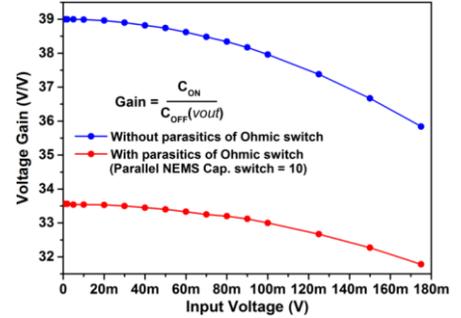

Fig.12. Variation of gain with respect to the input signal without parasitics of ohmic switch (*blue colour*) and with parasitics included (*red colour*).

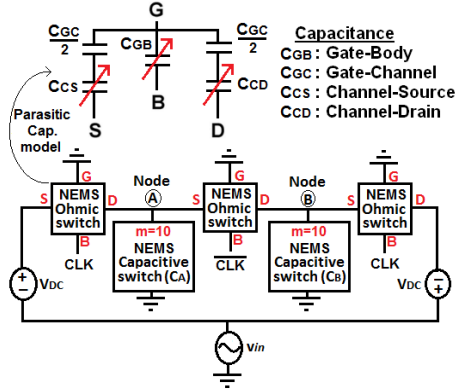

Fig.13. Modification to NEMS D-T amplifier for reducing the effect of parasitic capacitances of the ohmic switch.

Another major non-ideality that has to be accounted is the parasitic capacitances of the ohmic switch. Because, clock feed through and charge sharing at node "A" and "B" significantly affects the output voltage. In order to reduce its effect, the clock signal can be applied to the body terminal instead of gate and have multiple NEMS capacitive switch in parallel (say 10 parallel devices) instead of one, as shown in Fig. 13. In this way, the effect of parasitic capacitances can be reduced and the variation of gain in the modified circuit shown in Fig. 13 is plotted in Fig. 12 (*denoted in red color*).With the modified circuit as shown in Fig. 13, gain drops from its nominal value of 39 by 15% due to the charge sharing effect at nodes "A" and "B", which is evident from the gain plot in Fig.12 (*denoted in red color*).

*C. Power Dissipation*

The power dissipated in the NEMS DT-amplifier is only dynamic in nature. Since $v_{in}$, $V_{CLK} \ll V_{DC}$ and $C_{GB}$, $C_{GC} \ll (m \times C_A)$ the total power consumption of the amplifier is approximately given by $2 \times \{mC_A f_{CLK} V_{DC}^2\}$. Using the parameters of the amplifier discussed above, the total power consumed is 6.3 µW.

*D. Scaling of NEMS Capacitive switch*

As we saw in previous subsections, the main element which enables amplification is the NEMS capacitive switch (through its variable capacitance). For an amplifier to have higher dynamic range, higher $V_{PI}$ is required. Hence for lower dynamic range applications, the $V_{PI}$ can be reduced for lower power dissipation. Apart from the NEMS capacitive device used earlier, Table I shows two other prospective scaled versions of the device for different amplifier requirements. The C-V curve of the scaled versions of the device is plotted in Fig. 14.

TABLE I
DIFFERENT VERSIONS OF NEMS CAPACITIVE SWITCH

| Parameters of NEMS Capacitive switch | **High** voltage **High** gain | **Low** voltage **High** gain | **Low** voltage **Low** gain |
|---|---|---|---|
| | Large device[#] | Scaled version of device[+] | |
| L (µm) | 8.5 | 5 | 5 |
| W (µm) | 1.6 | 1 | 1 |
| t (nm) | 100 | 75 | 75 |
| Le (µm) | 7.9 | 4 | 4 |
| $g_o$ (nm)* | 135 | 50 | 50 |
| $t_d$ (nm)* | 27 | 10 | 20 |
| $\varepsilon_d$ * | 7.6 | 7.6 | 7.6 |
| $V_{PI}/V_{PO}$ | 9.6V / 6.2V | 3.8V / 2.4V | 4.0V / 2.7V |
| $C_{ON}/C_{OFF}$ | 31.5fF / 0.8fF | 26.9fF / 0.7fF | 13.5fF / 0.7fF |
| Voltage gain | 39 | 39 | 20 |

*Determines voltage gain; [#]used in previous sub-sections; [+]one possible scaling

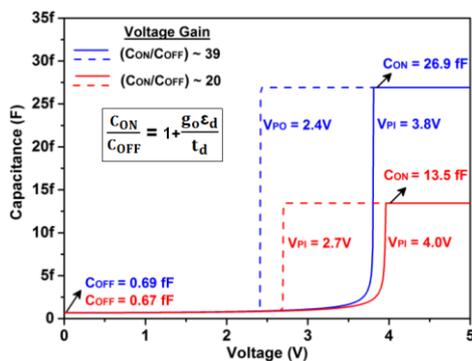

Fig.14. C-V curve of scaled low voltage NEMS capacitive switch. *Blue color* corresponds to high gain and *red color* corresponds to low gain application.

V. CONCLUSION

In this paper we have presented a new method of realizing a D-T amplifier using NEMS devices and is subsequently verified using the Verilog-A model. This opens up new application of utilizing NEMS devices for analog applications, instead of just being traditionally used as a passive switch. Even though lot of challenges lie ahead in realizing the above concept through fabricated devices, it provides a promising solution of reducing the energy consumption in ICs.